# Viewpoint: Opportunities and challenges of two-dimensional magnetic van der Waals materials: magnetic graphene?

Je-Geun Park[1,2*]

[1] Center for Correlated Electron Systems, Institute for Basic Science, Seoul 08826, Korea
[2] Department of Physics and Astronomy, Seoul National University, Seoul 08826, Korea
* Email: jgpark10@snu.ac.kr

**Abstract**: There has been a huge increase of interests in two-dimensional van der Waals materials over the past ten years or so with the conspicuous absence of one particular class of materials: magnetic van der Waals systems. In this Viewpoint, we point it out and illustrate how we might be able to benefit from exploring these so-far neglected materials.

A new material matters in condensed matter physics: so much so that if one looks at the history of the condensed matter physics over the last half-a century one immediately realizes that there have been star materials almost every decade. These then new classes of materials shaped and have been continuously shaping the big questions of the times. For example, there were spin glass, heavy fermion, high-temperature superconductors, manganites, and multiferroic materials, to name only a few.

The recent entry of graphene is no different from its predecessors, if not more revolutionary [1, 2]. One particularly amazing aspect of the graphene physics is how so much of the new physics was, at least initially, observed in graphene produced by such a simple mechanical exfoliation method using Scotch tape. This simple elegance of doing physics with Scotch tape in graphene impressed the community so much that it has ever since remained a method of first choice when it comes to producing monolayer van der Waals (vdW) materials. With the enormous recent developments, several new vdW materials have been discovered, or rediscovered to put it more correctly, and the list of the vdW materials is growing very fast [3].

However, there is one particular class of systems: magnetic vdW materials, conspicuously missing in the list known to date. In Figure 1, we show the numbers of annual publications on the subject of '*van der Waals*' as compared with those on the subject of '*magnetic, exfoliation*' for the past 15 years. Looking at the statistics, it is striking that although there were over 600 papers published in 2015 alone under the keyword of van der Waals materials there are less than 10 publications under the name of magnetic exfoliation. This statistics just amply illustrates how much the field of the magnetic van der Waals materials has been underexplored. As we argue in the remainder of this paper, however, we are in a very sorry state given the potential that would be made possible by this new class of materials. This paper, we hope, serves as a timely wake-up call for this situation.

Traditionally, vdW or layered magnetic materials have been considered as useful candidates for the study of low-dimensional magnetic systems [4]. Transition metal phosphorus trisuflide (or thiophosphate), $TMPS_3$, is one such example and its bulk properties have been extensively studied by using various techniques [5-11]. It is a very attractive aspect from a material's point of view that $TMPS_3$ can host several transition metal elements at the TM sites with correspondingly diverse physical properties: TM = Mn, Fe, Co, Ni, Zn, and Cd. One can also replace S by Se while keeping the same crystal structure, adding more flexibility to a choice of materials. This diversity in the materials with different physical properties will turn out to be a huge advantage when it comes to actual applications.

Interestingly enough, all three principal spin Hamiltonians are reported in these materials: two-dimensional (2D) Ising system ($FePS_3$), 2D Heisenberg system ($MnPS_3$), and 2D XY system ($NiPS_3$, $CoPS_3$) [8, 9]. From a study of the critical behavior, however $MnPS_3$ was claimed to be closer to an XY-like system [10]. More recently, it was theoretically proposed that a rare spin-valley coupling might be realized in one of these materials, $MnPS_3$ [12]. Another interesting theoretical finding is that upon carrier doping in the range of $10^{14}/cm^2$ the magnetic ground state can be tuned from the antiferromagnetic to ferromagnetic ground states [13]. Then $MnPS_3$ was reported to have a linear magnetoelectric (ME) coupling induced by the magnetic ordering and claimed to be a pure ferrotoroidic compound [14]. Adding further motivation to this class of materials, $TMPS_3$ has a band gap of 1.5 – 3.5 eV [15], nicely matching with the energy range of visible lights.

$TMPS_3$ has a weak van der Waals interaction between the layers and so can be easily cleavable. With hindsight, it is rather surprising to see why it has taken so long to produce a monolayer of these compounds. At least, now there are two independent reports for the realization of mono and a few layers of these systems [16-18]. It is interesting to note that the magnetic elements form a honeycomb lattice just like graphene so one can call it '*magnetic graphene*'.

Both groups have used the Scotch-tape method to achieve their goals and characterized their samples using the AFM and Raman techniques. As in graphene before, Raman spectroscopy is found to be a very useful tool in determining the thickness of TMPS$_3$: both E$_g$ and A$_{1g}$ Raman peaks show a clear thickness dependence [16]. The other noticeable exfoliated magnetic material is Bi-based high-temperature superconductor [3, 19]. With this successful mechanical exfoliation of TMPS$_3$, the door is now widely open to exploring the physical properties on the scale of few atomic layers and, more importantly, to exploiting its potentials for novel devices. We note that with these huge opportunities it is a welcome sign to see other magnetic vdW materials joining this rare group: for example, there have been reports on CrSiTe$_3$ [20].

One can think of several applications for the successfully exfoliated magnetic TMPS$_3$ monolayer. One of the most obvious cases is to use it for the study of fundamental 2D magnetism with reducing thickness as illustrated in Fig. 2. It sounds very strange, but to our best knowledge no experimental test has been done using a real magnetic material of the Onsager solution for 2D Ising magnets [21]. The only experimental test which we are aware of was carried out using sub-monolayer CH$_4$ absorbed on graphite, *an odd coincidence* [22]. It is not that we ever have any doubt about the answer that Onsager came up some 70 years ago, nevertheless it would be fantastic to see the results obtained from a real magnetic material confirming this historic achievement.

Moreover, with the variation of the magnetic atoms of TMPS$_3$ it will also be possible to extend this test of fundamental magnetism to other spin Hamiltonians and to demonstrate the Merin-Wagner-Hohenberg theorem [23, 24]. Another advantage of having the magnetic vdW monolayer of TMPS$_3$ is that Mott physics with strong correlation might be naturally realized in the 2D materials. If found correct, it will then open another window of fascinating opportunities to explore correlated physics on naturally occurring 2D systems. Furthermore, it will generally be an interesting question how other control parameters only available with the 2D magnetic systems: substrate and the width of the monolayer, affect the transition temperature. More specific to TMPS$_3$, it will also be intriguing to examine the strain effects on the magnetism as a recent theory suggested [25]. At the same time, with the band gap of TMPS$_3$ nicely overlapping with the energy range of visible lights, it will be interesting to investigate how the band gap varies as one reduces the thickness and/or the samples are put under external strain. The other potentially more far-reaching applications will be found in its use in conjunction with other vdW materials such as graphene and many other vdW materials as part of heterostructures. Over the past few years, we have witnessed an explosive growth of this field of vdW materials-based heterostructures with numerous novel discoveries ensuing therefrom [26, 27]. One is left only to guess how the already fast-growing field will change with the introduction of this new functionality of magnetism to the arsenal of vdW materials.

Despite the high notes, we have to admit that challenges lie ahead, in particular how to prepare the sample in a controller manner, e.g. with an accurate thickness control. The other problem that might hinder the progress is a lack of handy characterization tools. As most conventional techniques used for bulk magnetic materials have only limited usage for atomically thin magnetic vdW materials, we are in a desperate need of new techniques. However, with our own experience we can be sure that the following techniques will be found helpful for the field in future: Raman, AFM, PEEM, MFM, and MOKE.

All in all, we have no illusion as to how difficult and challenging the roads ahead will be even to achieve any one of the goals. However, failure or success will be another science-in-making in a true sense. Given the opportunities, it will certainly be worth taking.

We acknowledge Cheol-Hwan Park for his critical reading of the manuscript and suggestions. The work at the IBS CCES was supported by the research program of Institute for Basic Science (IBS-R009-G1)

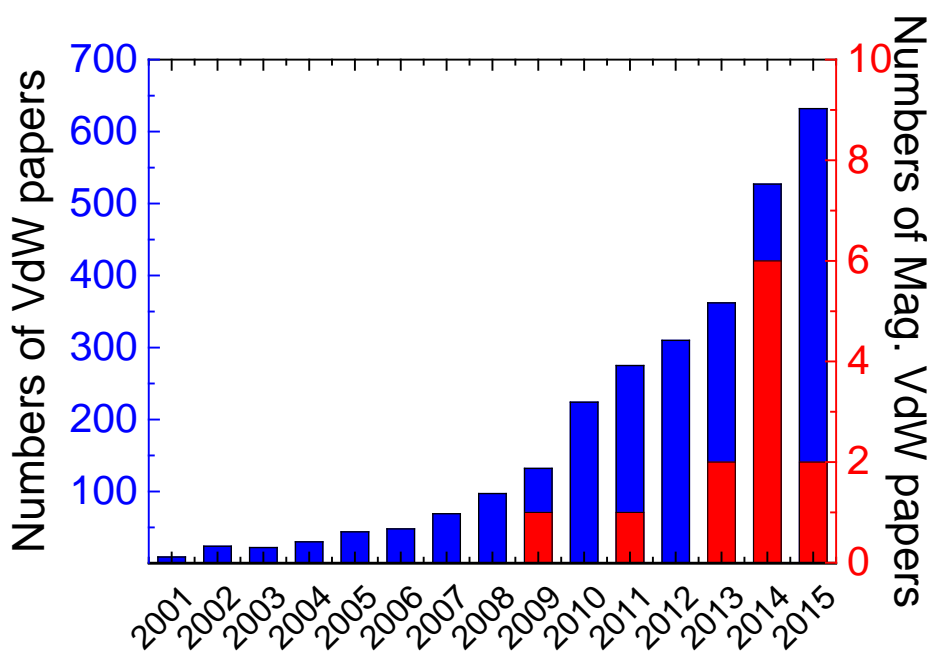

**Figure 1** Number of annual publications over the past 15 years with the key words of (left axis & blue colour) 'van der Waals (vdW)' and (right axis & red colour) 'magnetic exfoliation (magnetic vdW)'. We used the Google scholar search engine to collect the statistics.

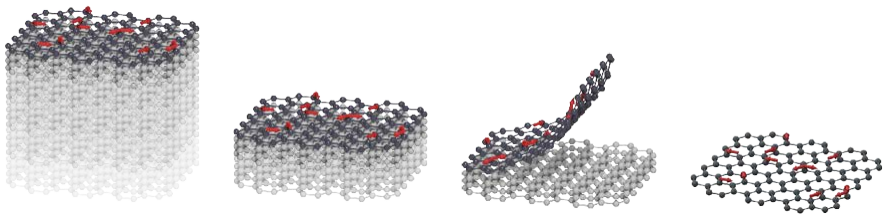

**Figure 2** Schematic of how one can study the thickness dependence of the fundamental magnetic properties using the magnetic vdW materials such as the transition temperatures and the ground states.